\documentclass[prx,sd,amsmath,amssymb,reprint]{revtex4-1}

\usepackage{graphicx}% Include figure files
\usepackage{bm}% bold math
\usepackage{color}

\begin{document}

%\preprint{APS/123-QED}

\title{Mapping the metastability of Lennard-Jones clusters by the maximum vibrational frequency}

\author{Shota Ono}
\email{shota\_o@gifu-u.ac.jp}
\affiliation{Department of Electrical, Electronic and Computer Engineering, Gifu University, Gifu 501-1193, Japan}

\begin{abstract} 
We study the structure-stability relationship of the Lennard-Jones (LJ) clusters from a point of view of vibrations. By assuming the size up to $N=1610$, we demonstrate that the $N$-dependence of the maximum vibrational frequency reflects the geometry of the core (the interior of cluster) that will determine the overall geometry of the cluster. This allows us to identify the formation of non-icosahedral structures for $N\le 150$, the vacancy formation at the core for $N\ge 752$, and the transition from icosahedral to decahedral structures at $N = 1034$. We apply the maximum frequency analysis to classify metastable clusters for $19\le N \le 39$, where transformation pathways between different structures are visualized, and the energy barrier height is estimated simultaneously.   
\end{abstract}
%\pacs{}

\maketitle

%%%%%%%%%%%%%%%%%%%%%%%%%%%%%%%%%%%%%%%%%%%%%%%%%%%
\section{Introduction}
Assembly of atoms constitutes nanoclusters, and the structure, thermodynamics, and growth process have been extensively investigated for many elemental systems such as Pb \cite{lead2005,lead2006,lead2011}, Ni \cite{ni11}, Au \cite{au1,au2}, and transition metals \cite{3d4d5d,nano2017} (see also Ref.~\cite{NP} for a review). The number of atoms $N$ plays an important role in understanding the stability of clusters because the total energy at a specific $N$ is particularly low, compared to that at $N\pm 1$. In addition, the cluster geometry can be highly symmetric. For example, the clusters at $N=13$ and $55$ can have relatively small energy, which is usually attributed to the icosahedral geometry. On the other hand, for large $N$, the structure of clusters is determined by several factors (i.e., the volume, surface, edges, and vertices of clusters), and the geometry changes from the icosahedral to decahedral to face-centered cubic (fcc) structures as $N$ increases \cite{NP}: For example, the clusters with the decahedral structure have been recently created for several noble metals \cite{zhou}, which has attracted attention due to their optical and catalytic properties that are different from those with the icosahedral structure. 

The stability and geometry of the Lennard-Jones (LJ) clusters have been extensively studied for many years \cite{northby,raoult,xue,deaven,wales1997,leary,doye1,doye2,solo,polak2003,shao310-561,shao2004,shao_vac,shao2005,noya,yang}. It has been known that the $N=13$ icosahedron serves as a seed to generate the lowest energy atomic configurations \cite{solo}, that is, the LJ clusters for $N\ge 13$ have an icosahedron at the core surrounded by the surface atoms. However, for the cases of $N=$ 38, 75-77, 98, and 102-104, the core of the LJ clusters has octahedral, decahedral, tetrahedral, and decahedral structures, respectively \cite{wales1997,leary,solo}. Even when the energetic stability analyses are employed, no significant anomalies have been found at these $N$s. In general, the cluster geometry is characterized by studying the local atomic environment in detail, as done by Polak and Patrykiejew \cite{polak2003} and Yang and Tang \cite{yang}, where they used four structural motifs (fcc, hcp, icosahedral, and decahedral) to understand the overall and core geometries. Alternatively, we expect that the lattice dynamics calculations might be useful to understand the core geometry because the normal modes at the maximum frequency will involve the vibration of the most rigid part in the system. For example, the sequence of the maximum frequency as a function of $N$ should identify the difference of the core geometries in the LJ clusters.

In this paper, we study the energetic and vibrational properties of LJ clusters up to $N=1610$. The $N$-dependence of the maximum frequency allows us to identify the core geometry that is different from the icosahedral structure. It also enables us to identify the vacancy formation at the core and the structural transition from icosahedra to decahedra for large $N$. As another application, we construct a metastability map, where the maximum frequency is plotted as the total energy for many metastable structures. For the cases of $19 \le N \le 39$, we distinguish the clusters with decahedra from those with icosahedra. In addition, we identify transformation pathways between different structures, and estimate the energy barrier height. The present work will pave the way to understand the structural stability and geometry based on the vibrational frequency. 

%The standard deviation of the one-particle energy also reflects the core geometry of LJ clusters, where the sum of the one-particle energy is equal to the total energy of the system.

The vibrational frequency analysis has been recently applied to study the magic numbers in $N$ charges on a sphere \cite{ono_magic}, while finding the lowest energy configurations on a sphere is known as the Thomson problem. The maximum frequency showed relatively small values at $N=$ 12, 32, 72, 132, 192, 212, 272, 282, and 372. The presence of these magic numbers reflects both the charge configurations on a sphere and the strong degeneracy of the one-particle energies. In contrast, the LJ particles that we study in the present work are free from the boundary condition, and therefore the core geometry of the system influences the vibrational properties. Doye and Calvo have calculated the geometric mean vibrational frequency of the LJ clusters to distinguish the non-icosahedral structures from the icosahedral structure at the selected sizes of $N=38$, 75, 98, and 102 \cite{doye2}. The maximum frequency that we use in the present work is directly related to the vibration of the core around which the interatomic bonding is the strongest in the cluster. 

Our approach can provide a useful insight of the structure-stability relationship among metastable structures as well, only by performing the maximum frequency and total energy calculations. The metastability of nanoclusters has been studied by calculating the potential energy surface (PES) and/or constructing the disconnectivity graph \cite{becker}. These approaches are useful to understand the relationship between the energetic stability and the structure of nanoclusters, which has been applied to predict the synthesizability of nanostructures \cite{wales1998,de}. However, visualizing the PES is not straightforward \cite{shires}: among $3N$ degrees of freedom one must find a few parameters describing the transformation between different structures. To construct the disconnectivity graph, one must find the transition states having one imaginary frequency mode. 

%It is of primary importance to find a seed or a building block in understanding the stability of condensed matters because a wide variety of structures can be systematically constructed from the fundamental units.  

%%%%%%%%%%%%%%%%%
\begin{figure*}[t]
\center
\includegraphics[scale=0.45]{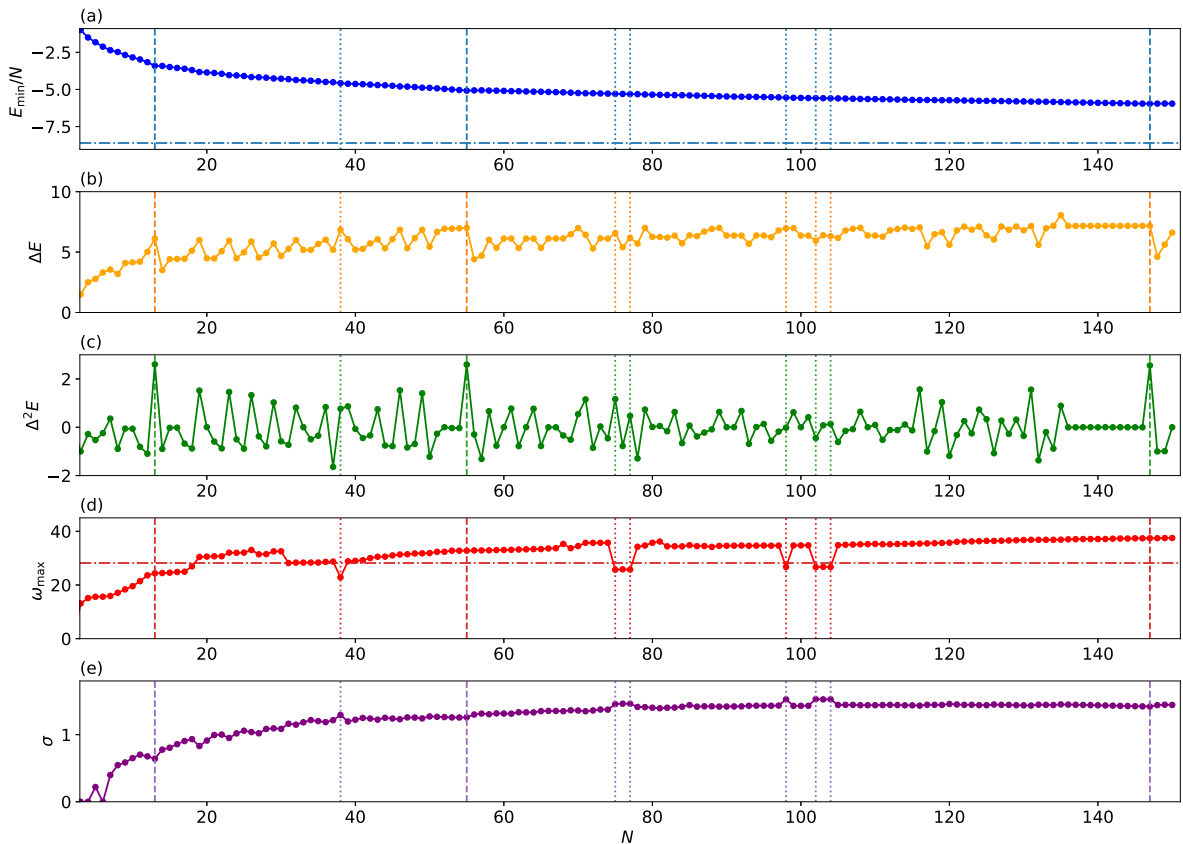}
\caption{The $N$-dependence of (a) $E_{\rm min}/N$, (b) $\Delta E$, (c) $\Delta^2 E$, (d) $\omega_{\rm max}$, and (e) $\sigma$. The vertical dashed lines indicate $N=13, 55,$ and $147$ (complete icosahedral structure). The vertical dotted lines indicate $N=38, 75, 77, 98, 102, $ and $104$ (non-icosahedral structure). The horizontal dot-dashed lines in (a) and (d) indicate the values for the fcc structure. } \label{fig_1} 
\end{figure*}
%%%%%%%%%%%%%%%%%

%%%%%%%%%%%%%%%%%%%%%%%%%%%%%%%%%%%%%%%%%%%%%%%%%%%
\section{Theory}
\label{sec:theory}
%%%%%%%%%%%%%%%%%%%%%%%%%%%%%%%%%%%%%%%%%%%%%%%%%%%

We study the dynamics of the LJ clusters having $N$ atoms within the harmonic approximation. The equations of motion can be written as \cite{ziman,mermin}
\begin{eqnarray}
m \frac{d^2 u_{i\alpha}}{dt^2}  
 = - \sum_{j\beta} D_{\alpha\beta}^{ij} u_{j\beta},
\end{eqnarray}
where $u_{i\alpha}$ is the displacement along $\alpha$ direction for the particle $i$ with a mass of $m$. The force constant matrix $D_{\alpha\beta}^{ij}$ is defined as 
\begin{eqnarray}
D_{\alpha\beta}^{ij} = D_{\beta\alpha}^{ji} =
 \frac{\partial^2 E}{\partial R_{i\alpha} \partial R_{j\beta}} \Big\vert_0,
 \label{eq:dynmat}
\end{eqnarray}
where the derivative is taken at the equilibrium configurations. $E$ is the total potential energy 
\begin{eqnarray}
 E &=& \sum_{i=1}^{N} \varepsilon_i 
 \label{eq:Vpot}
\end{eqnarray}
with the one-particle energy 
\begin{eqnarray}
\varepsilon_i &=& \frac{1}{2} \sum_{j\ne i} 4A \left[ \left( \frac{\sigma}{r_{ij}} \right)^{12} - \left(\frac{\sigma}{r_{ij}} \right)^{6} \right],
\label{eq:epsi}
\end{eqnarray}
where $1/2$ accounts for the double counting of the interaction energy, $A$ and $\sigma$ are parameters of the LJ potential, and $r_{ij}$ is the interparticle distance between the LJ particles $i$ and $j$, which can be expressed by 
\begin{eqnarray}
 r_{ij}^2 = \sum_{\alpha=x,y,z} (R_{i\alpha}-R_{j\alpha})^2,
\end{eqnarray}
where $R_{i\alpha}$ is the $\alpha$ component of the position of the particle $i$. Assuming a stationary solution $u_{i\alpha}(t) = \epsilon_{i\alpha} e^{i\omega t}$ with the frequency $\omega$ and the polarization $\epsilon_{i\alpha}$, one obtains the eigenvalue equation
\begin{eqnarray}
m \omega^2 \epsilon_{i\alpha} 
&=& \sum_{j\beta}
D_{\alpha\beta}^{ij} \epsilon_{j\beta}.
\label{eq:eigen}
\end{eqnarray}
The stable structure with $N$ particles has $3N-6$ vibrational modes, where the degrees of freedom for translation and rotation are subtracted. The maximum eigenvalue gives the maximum frequency $\omega_{\rm max}$. The units of energy and frequency are $A$ and $A^{1/2}\sigma^{-1}m^{-1/2}$, respectively. Throughout the paper, we set $A=\sigma=m=1$. 

To find the lowest energy structures, we referred to two databases: For $3\le N\le 150$, we referred to the Cambridge Cluster Database (CCD) \cite{CCD}, and for large $N$ up to 1610, we referred to the database provided by Shao {\it et al.} \cite{shaoCCD} (310 to 561 atoms \cite{shao310-561}, 562 to 1000 atoms \cite{shao2004}, and 1001 to 1610 atoms \cite{shao2005}). We used the Broyden-Fletcher-Goldfarb-Shanno algorithm \cite{numerical_recipe} to find the local minimum structures for the cases of $N=19$-39, where the initial positions of $N$ atoms were given by random numbers. For each $N$, we generated more than $3\times 10^4$ initial configurations and optimized their structures, from which the 2000 lowest energy structures were extracted. In particular, for $N=31$, 33, and 35, $6\times 10^4$ initial configurations were needed to obtain the lowest energy structures stored at the CCD \cite{CCD}. However, we failed to find the lowest energy structure at $N=38$. 

To compare the optimized $E/N$ and $\omega_{\rm max}$ with the bulk values (i.e., the case of $N\rightarrow \infty$), we calculated the total energy and the phonon dispersions of the LJ crystal in the fcc structure. The computational details are the same as those described in Ref.~\cite{ono_ito}, while $A$ and $\sigma$ were set to be unity in the present work. We obtained $E_{\rm min}^{\rm fcc}/N=-8.609$ and $\omega_{\rm max}^{\rm fcc}=28.18$ that corresponds to the longitudinal phonon frequency at the $X$ point in the Brillouin zone. 

%Negative eigenvalues (i.e., imaginary frequencies) indicate that the configuration obtained is unstable against the corresponding vibration mode. We then displace the particles $i$ along the direction of the eigenvector and reoptimize $V$. This procedure was performed until no negative eigenvalues were obtained, which will provide the ground state or metastable states. 

%%%%%%%%%%%%%%%%%
\begin{figure}[t]
\center
\includegraphics[scale=0.42]{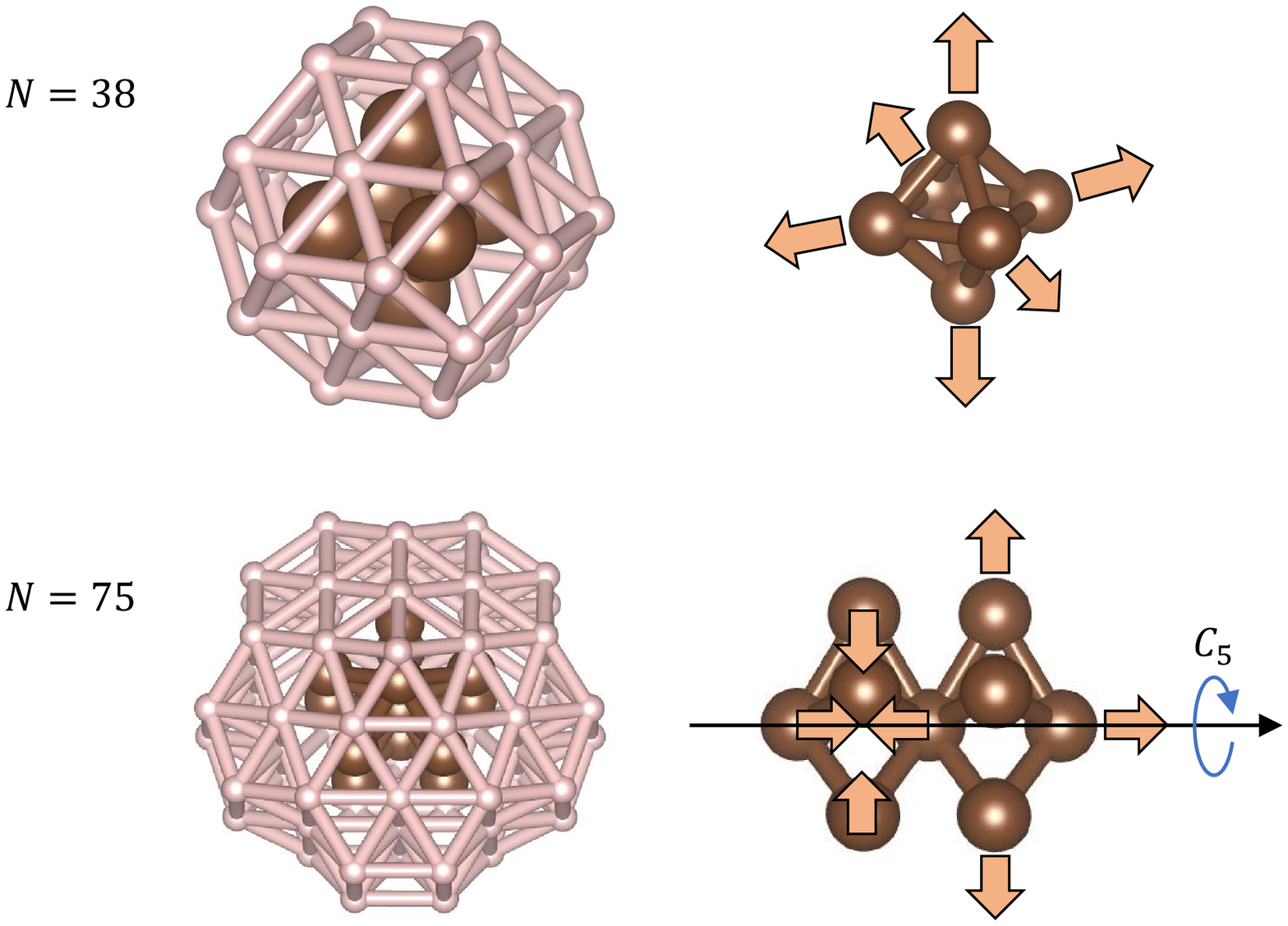}
\caption{The structure of the lowest energy atomic configurations (left) and the displacement vectors of the maximum frequency mode (right) for $N=38$ and 75. The atoms at the core region are colored brown. The core at $N=75$ has the five-fold rotational symmetry. } \label{fig_2} 
\end{figure}
%%%%%%%%%%%%%%%%%

%%%%%%%%%%%%%%%%%
\begin{figure}[t]
\center
\includegraphics[scale=0.45]{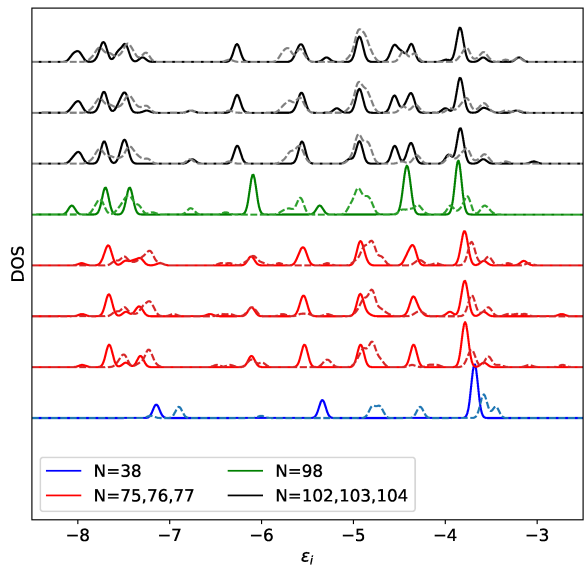}
\caption{The one-particle DOS for $N=$ 38, 75, 76, 77, 98, 102, 103, and 104 clusters in the lowest energy structure (solid) and in the icosahedral structure (dashed). } \label{fig_3} 
\end{figure}
%%%%%%%%%%%%%%%%%

%%%%%%%%%%%%%%%%%
\begin{figure*}
\center
\includegraphics[scale=0.5]{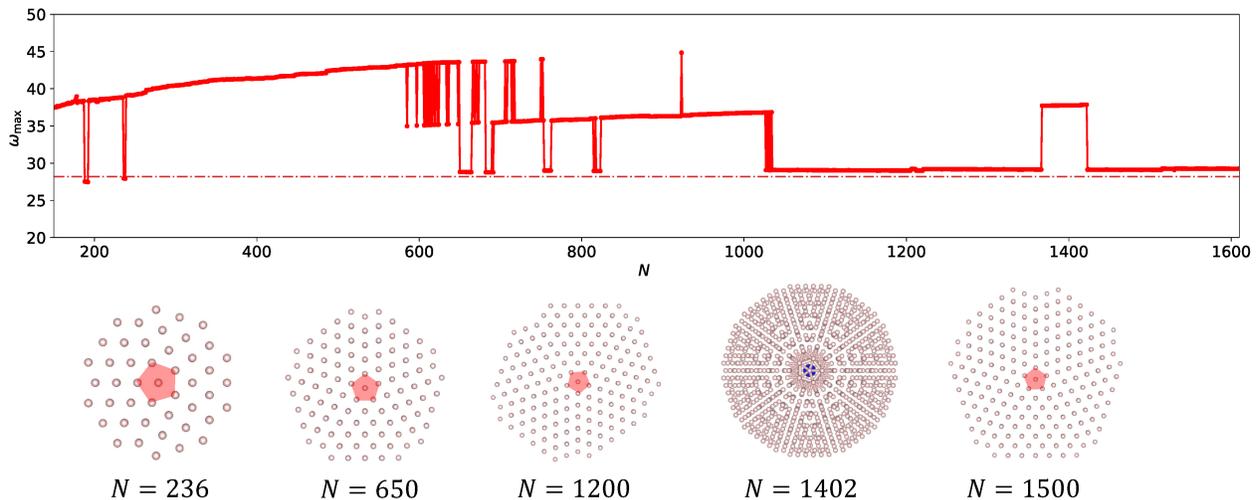}
\caption{The $N$-dependence of $\omega_{\rm max}$ for the lowest energy structures: $151\le N \le 1610$. The dot-dashed line indicates the $\omega_{\rm max}^{\rm fcc}$. The four clusters with the decahedral structure (colored red) are shown. The $N=1402$ cluster with the icosahedral structure is also shown, where the central vacancy is colored blue and the interatomic bonding is illustrated for $r_{ij}\le 1.05\sigma$. } \label{fig_4} 
\end{figure*}
%%%%%%%%%%%%%%%%%

%%%%%%%%%%%%%%%%%%
\begin{table*}
\begin{center}
\caption{The values of $\varepsilon_i$ with $i=1$ to $7$ for several $N$s. The figure in parenthesis indicates the degeneracy. The ``38g'' and ``38i'' indicate the global minimum and the icosahedral structure at $N=38$, respectively. }
{
\begin{tabular}{lccccccccc} \hline\hline
$N$ & $\varepsilon_1$ & $\varepsilon_2$ & $\varepsilon_3$ & $\varepsilon_4$ & $\varepsilon_5$ & $\varepsilon_6$ & $\varepsilon_7$ & $\varepsilon_8$ & $\varepsilon_9$ \\
\hline
38g \hspace{0mm} & $-7.146$ (6) \hspace{0mm} & $-5.338$ (8) \hspace{0mm} & $-3.681$ (24) \hspace{0mm} & - \hspace{0mm} & - \hspace{0mm} & - \hspace{0mm} & - & -  \\
38i \hspace{0mm} & $-7.208$ (1) \hspace{0mm} & $-6.900$ (5) \hspace{0mm} & $-6.003$ (1) \hspace{0mm} & $-4.795$ (5) \hspace{0mm} & $-4.713$ (5) \hspace{0mm} & $-4.274$ (5)  \hspace{0mm} & $-3.584$ (10) \hspace{0mm} & $-3.540$ (1) \hspace{0mm} & $-3.450$ (5) \\ 
\hline\hline
\end{tabular}
}
\label{table_1}
\end{center}
\end{table*}
%%%%%%%%%%%%%%%%%
%%%%%%%%%%%%%%%%%
\begin{figure*}
\center
\includegraphics[scale=0.5]{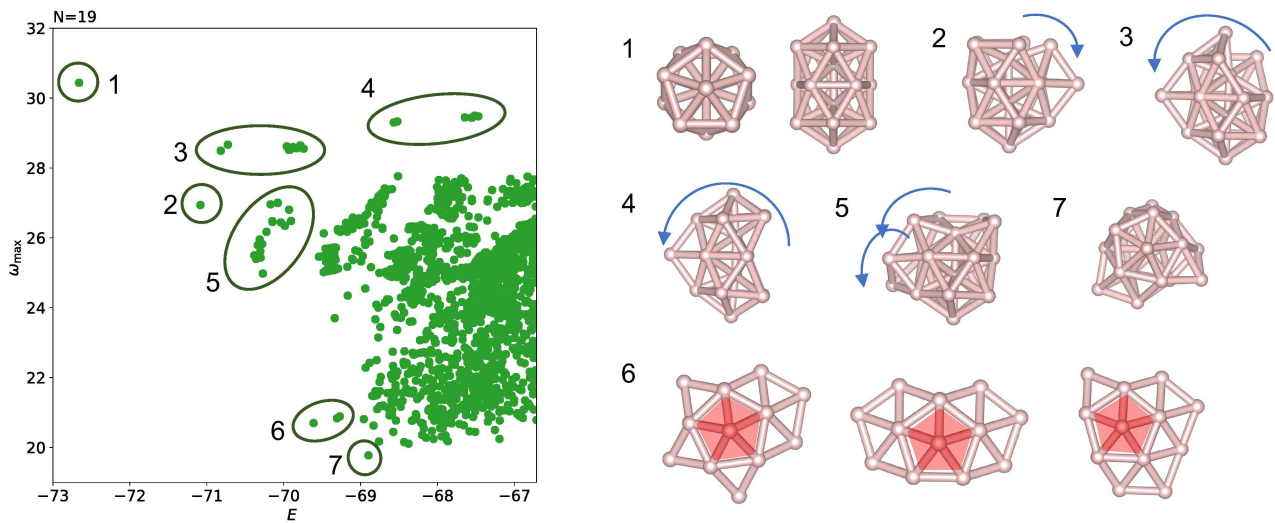}
\caption{(left) The distribution of $\omega_{\rm max}$ versus the total energy of $E$ for the 2000 lowest energy structures with $N=19$. The plotted points are classed into seven groups, where the group 6 includes three points. (right) The illustration of metastable structures of groups 1-7, where the group 6 structures have a decahedron (colored red). The arrows indicate the atomic movements from the group 1 structure. } \label{fig_5} 
\end{figure*}
%%%%%%%%%%%%%%%%%

%%%%%%%%%%%%%%%%%
\begin{figure*}
\center
\includegraphics[scale=0.55]{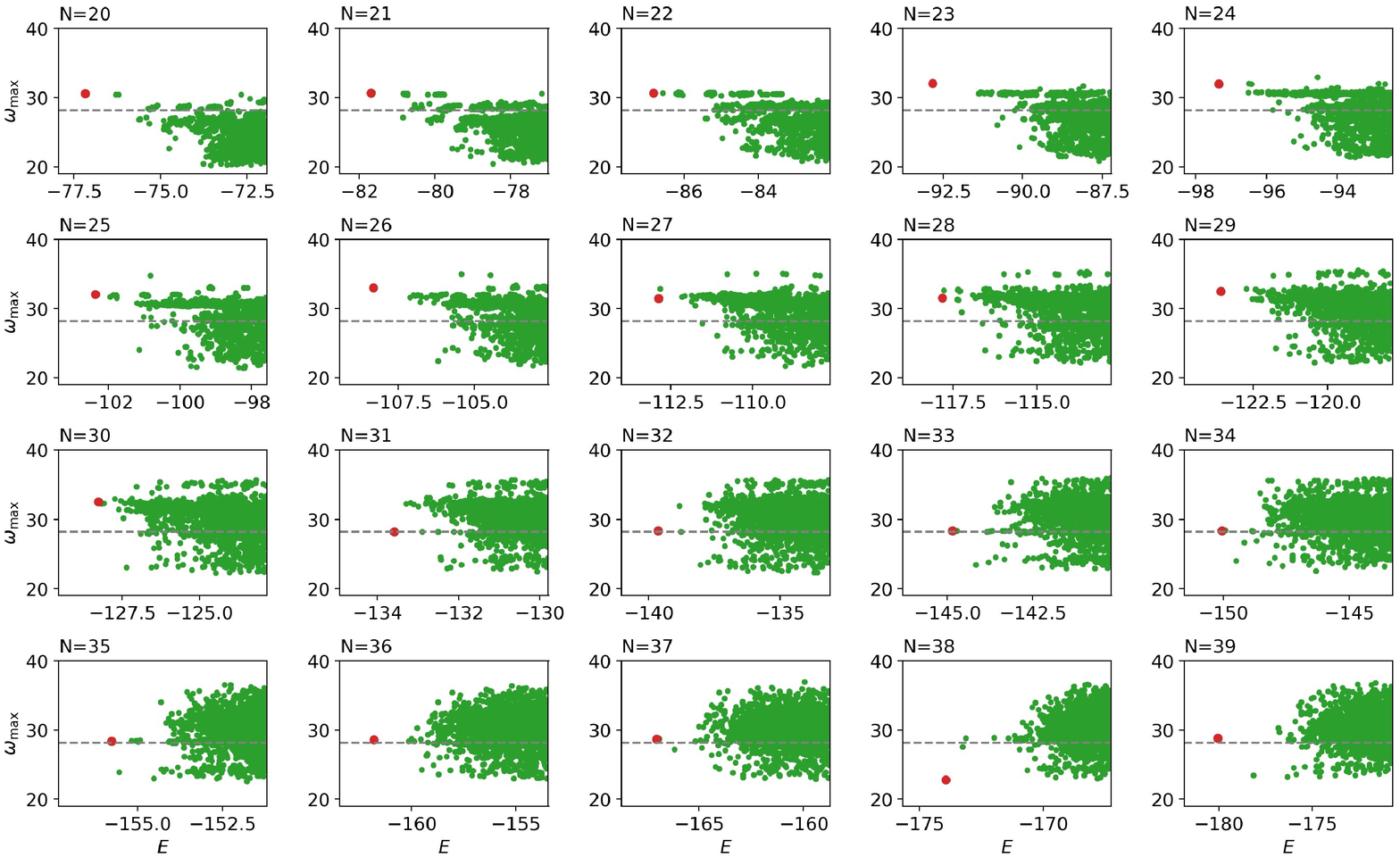}
\caption{The distribution of $\omega_{\rm max}$ versus the total energy of $E$ for the 2000 lowest energy structures with $N=20$-39. The data of the lowest energy structure is indicated by a large circle (colored red). The dashed line indicates the $\omega_{\rm max}^{\rm fcc}$.} \label{fig_6} 
\end{figure*}
%%%%%%%%%%%%%%%%%

%%%%%%%%%%%%%%%%%%%%%%%%%%%%%%%%%%%
\section{Results and Discussion}
\subsection{Lowest energy structures}
Figure \ref{fig_1}(a) shows the $N$-dependence of the lowest energy per particle $E_{\rm min}/N$. The $E_{\rm min}$ decreases with $N$, while some dips can be observed at $N=13$ and $N=55$. At these $N$s complete icosahedral structure can be formed \cite{deaven,solo}. To emphasize the magic numbers, at which the $N$ cluster is relatively stable compared to $N\pm 1$ clusters, we define the first and second differences in the total energy as \cite{NP}
\begin{eqnarray}
 \Delta E &=& E_{\rm min}(N-1)-E_{\rm min}(N),
 \\
 \Delta^2 E &=& E_{\rm min}(N-1)+E_{\rm min}(N+1)-2E_{\rm min}(N).
\end{eqnarray} 
Figures \ref{fig_1}(b) and \ref{fig_1}(c) show the $N$-dependence of $\Delta E$ and $\Delta^2 E$, respectively. Peaks in $\Delta^2 E$ are observed at several $N$s. For example, $N=13, 55,$ and $144$ are clearly identified as the magic numbers, where $N=144$ cluster has also complete icosahedral structure. The relative stability between different $N$s can thus be identified by studying $\Delta E$ and/or $\Delta^2 E$. 

To shed light on another aspect on the LJ cluster properties, we show the $N$-dependence of $\omega_{\rm max}$ in Fig.~\ref{fig_1}(d). The value of $\omega_{\rm max}$ increases with $N$. However, an anomalous decrease in $\omega_{\rm max}$ is observed at $N=38, 75$-77, 98, and 102-104. It should be noted that the lowest energy structures at these $N$s do not have icosahedra at the core \cite{wales1997,leary,solo,CCD}. This shows that the atomic displacement of the maximum frequency mode reflects the core geometry: For example, $N=38$ and 75 clusters has an octahedron and a decahedron at the core, and the displacement pattern is a breathing of the octahedron for $N=38$ and asymmetric displacements along the five-fold symmetric axis for $N=75$, as shown in Fig.~\ref{fig_2}. The maximum frequency mode for the $N=76, 77$, and 102-104 clusters is similar to that for the $N=75$ cluster: the core at $N=$ 102-104 has 19 atoms forming a one-dimensional tube of three decahedra with the five-fold rotational symmetry, and the displacement pattern shows the expansion, contraction, and expansion of the three decahedra. The $N=98$ cluster has a large core with tetrahedral shape \cite{leary}, and the displacement localizes to this core in a complicated manner. In this way, the analysis of the $N$-dependence of $\omega_{\rm max}$ allows us to distinguish the core geometry from icosahedra. 

The difference of the core geometry as well as the decrease in $\omega_{\rm max}$ are related to the distribution of $\varepsilon_i$ in Eq.~(\ref{eq:epsi}). In general, the $\varepsilon_i$ of the core atom $i$ is lower than that of the surface atoms because the interatomic bonding strength as well as the coordination number will be large around the core. The values of $\varepsilon_i$ are thus scattered as $N$ increases because the core and surface regions are clearly separated for large $N$. The distribution of $\varepsilon_i$ will be modified when the geometry of the core changes. Figure~\ref{fig_1}(e) shows the standard deviation $\sigma$ of $\varepsilon_i$ as a function of $N$. The peaks in $\sigma$ are well correlated to the decrease in $\omega_{\rm max}$ in Fig.~\ref{fig_1}(d).  

The increase in $\sigma$ can be visualized by the density-of-states (DOS) for $\varepsilon_i$ shown in Fig.~\ref{fig_3}, where the DOS of the lowest energy (non-icosahedral) structure and the local minimum (icosahedral) structure (extracted from the CCD \cite{CCD}) are shown for $N=38, 75$-77, 98, and 102-104. The gaussian broadening method is applied, and the DOS is shifted depending on the $N$. The DOS peaks observed for the lowest energy structure are smeared out or split when the icosahedral structure (a local minima) is considered. The value of $\varepsilon_i$ and the degeneracy for the $N=38$ are listed in Table \ref{table_1}. 

As shown in Fig.~\ref{fig_1}(a), the $E_{\rm min}/N$ approaches the $E_{\rm min}^{\rm fcc}/N$ as $N$ increases. However, the energy difference is still large: $E_{\rm min}/N=-5.955$ for $N=150$, and the relative error is more than 30 \%. When $N$ is increased up to 1610, $E_{\rm min}/N=-7.338$ \cite{shao2005}, and the error is reduced to 15 \%. It is interesting to study how the $\omega_{\rm max}$ approaches the bulk value of $\omega_{\rm max}^{\rm fcc}$ because the $N$-dependence of $\omega_{\rm max}$ is nonmonotonic for small $N$, as shown in Fig.~\ref{fig_1}(d). In addition, Shao {\it et al}. have proposed the vacancy formation at the core for $N\ge 752$ \cite{shao_vac} and the structural transition from icosahedra to decahedra at $N=1034$ \cite{shao2005}, which will influence the $N$-dependence of $\omega_{\rm max}$. Figure \ref{fig_4} shows the $\omega_{\rm max}$ as a function of $N$ up to 1610: (i) The $\omega_{\rm max}$ increases from 37 to 43 for $151\le N \lesssim 600$, except for $188 \le N \le 192$ and $236 \le N \le 238$; (ii) the value of $\omega_{\rm max}$ decreases to 35 around $N\simeq 600$, deviates around 35 for $600 \lesssim N \lesssim 800$, and increases from 35 to 36 for $800 \lesssim N \le 1034$; and (iii) the $\omega_{\rm max}$ shows a sudden drop to 29 at $N=1035$, and almost keeps the constant value of 29 up to $N =1610$, while the jump within $1367 \le N \le 1422$ is observed. 

The property (i), an increase in $\omega_{\rm max}$ with $N$, indicates the hardening of the core in the cluster, which is also observed in small $N$ (see Fig.~\ref{fig_1}(d)). The property (ii), an significant decrease in $\omega_{\rm max}$ from 43 to 35, reflects the vacancy formation at the core: for $N\ge 752$ the icosahedral structures with the central vacancy are more stable except for $N=923$ \cite{shao_vac}. We consider that some of the peaks around $600 \lesssim N \lesssim 800$ and $N=923$ will be a reminiscence of the behavior for $N\lesssim 600$. The property (iii) indicates the transition from the icosahedral to decahedral structures \cite{shao2005}. The jump of $\omega_{\rm max}$ within $1367 \le N \le 1422$ is also consistent with the formation of the icosahedral structure with the central vacancy at $N=1402$ \cite{shao2005}. In this sense, the anomalous decreases around $188 \le N \le 192$, $236 \le N \le 238$, $650 \le N \le 664$, $682 \le N \le 689$, $755 \le N \le 762$, and $815 \le N \le 823$ can be attributed to the formation of the decahedral structure. The decahedral and icosahedral structures for the selected $N$s are shown in Fig.~\ref{fig_4}. 

It should be noted that the $N$-dependence of $\omega_{\rm max}$ is strongly correlated with that of the fcc motif concentration. Yang and Tang studied how many structural motifs are there in the LJ cluster for each $N$ by considering four types of motifs in the fcc, hcp, icosahedral, and decahedral structures with 13 atoms, and showed that the fcc motif concentration is relatively large when $N=38$, 75-78, 102-104, 188-192, 236-238, and $N\simeq 1030$ \cite{yang}. They attributed such an enhancement to the formation of the Marks decahedron rather than the Mackay icosahedron, which is consistent with our characterization in geometry. 

%%%%%%%%%%%%%%%%%%%%%%%%%%%%%%%%%%%%%
\subsection{Metastable structures}
We next apply the $\omega_{\rm max}$-based analysis to a classification of the metastable structures with $N$ fixed. Figure \ref{fig_5} shows the distribution of $\omega_{\rm max}$ as a function of $E$ for the case of $N=19$, where the 2000 lowest energy structures are plotted. Data points form an island in the $E$-$\omega_{\rm max}$ plane, which enables us to classify metastable structures into some groups. The lowest energy structure (group 1) is a barrel-shaped double icosahedron, where four atoms form the symmetry axis along which three pentagons are stacked with twisted angle of $\pi/5$. The structures of other groups are basically derived from the group 1 structure (see Fig.~\ref{fig_5}): in the second lowest energy structure (group 2), the vertex located on the symmetry axis moves to another facet; in the group 3 and 4 structures, the vertex on the first (or third) and second layer of the pentagon moves to another facet, respectively; and in the group 5 structures, two vertices moves to other facets. As one approaches the continent (i.e., densely plotted region for $E\ge -69$), the structures with more complex geometry are observed.

One can find three anomalous structures, which are apart from the continent, with $-70 \le E \le -69$ and relatively small values of $\omega_{\rm max}$ ($\simeq 20.8$). As depicted in Fig.~\ref{fig_5}, those metastable structures (group 6) have a decahedron at the core. This result also confirms that the clusters with non-icosahedra can have relatively small $\omega_{\rm max}$. 

%This is an opposite tendency when the mean frequencies are plotted as the total energies \cite{doye2}. 
%the 34, 89, and 94th lowest energy structures, the 133th, the 198th

We also found that some of the structures have a decahedron: For example, those structures have $(E, \omega_{\rm max})=$ $(-68.9578, 20.81)$ and $(-68.6245, 20.92)$. Among the 2000 structures generated in the present calculations, the 142th structure ($E = -68.8976$) has the lowest maximum frequency of $\omega_{\rm max}=19.77$, but has neither a decahedron nor an icosahedron at the core region (see group 7 in Fig.~\ref{fig_5}). 

The $E$-$\omega_{\rm max}$ map gives a rough estimation of the energy barrier height between different structures. For example, the transformation from the group 1 to group 2 can be possible when the energy about $2A$ is added. On the other hand, to obtain the clusters with the decahedral structures, the structure in the group 1 must first move to the continent around $E\ge -68$, and next moves to the island of the group 6, implying that more than $5A$ (rather than $3A$) is needed. This view is consistent with that obtained by the disconnectivity graph calculations \cite{doye1}. 

%This behavior is quite similar to the disconnectivity graph that has been frequently used to study the metastability analysis of finite size clusters by calculating the transition states. The implementation of the present approach is easy because one only needs to calculate the total energy and the maximum frequency of the optimized structures.

It is interesting that when the metastable structures differ strongly in geometry, those are located at different region in the $E$-$\omega_{\rm max}$ plane, as shown in Fig.~\ref{fig_5}. This allows us to study the evolution of the cluster geometry with the size $N$. Figure \ref{fig_6} shows how the distribution of $\omega_{\rm max}$ evolves within the range of $20\le N \le 39$. When $20\le N \le 30$, the distribution of $\omega_{\rm max}$ is similar to that for the case of $N=19$: the lowest energy structure has a relatively high value of $\omega_{\rm max}$, while some metastable structures have relatively small values of $\omega_{\rm max}\simeq 22$. The former structures are constructed by adding atoms to the surface region of the lowest energy structure of $N=19$, i.e., a double icosahedron, whereas the latter structures consist of decahedra. When $N$ is increased to 31, the $\omega_{\rm max}$ of the lowest energy structure becomes small ($\omega_{\rm max}\simeq 28$), implying that the number of decahedra overcomes that of icosahedra. For $N\ge 32$, a strong distribution around $\omega_{\rm max}=32$ is smeared out, and the distribution of $\omega_{\rm max}$ tends to be symmetric around $\omega_{\rm max}=30$. In the $E$-$\omega_{\rm max}$ plane, islands evolve for low $\omega_{\rm max}$ when $N$ is increased, producing the octahedral structure at $N=38$. For $N=39$, the lowest energy structure ($\omega_{\rm max}=28.8$) is constructed from a decahedron surrounded by five decahedra, whereas the second lowest energy structure ($\omega_{\rm max}=23.5$) is constructed by adding an atom to the $N=38$ octahedral structure. The energy barrier hight between them is estimated to be more than $5A$ rather than $2A$. In this way, the construction of the $E$-$\omega_{\rm max}$ map is a useful method to understand the transformation between metastable structures, which will be alternative to the disconnectivity graph constructions \cite{doye1}. 

%This might be a precursor of the transformation from the decahedral to the fcc structures. 

The structural transition from the icosahedral to decahedral structures has been discussed in the literature. Deaven {\it et al}. identified a significant change in the one-particle energy distribution \cite{deaven}: The peak of the lowest one-particle energy shifts dramatically from $\varepsilon_i \simeq -6.5$ at $N=30$ to $\varepsilon_i \simeq -7$ at $N=31$. On the other hand, Raoult {\it et al}. \cite{raoult} and Shao {\it et al}. \cite{shao2005} showed that the structural transition to the decahedral structure occurs at $N>1000$ by performing the the total energy calculations. The present calculations suggest that the profile of the $\omega_{\rm max}$ distribution is quite different across $N=31$. We expect that the systematic calculations of the $\omega_{\rm max}$ distribution for large $N$ enable us to understand the structural transitions (from the icosahedral to decahedral, and from the decahedral to fcc structures as well). 

%For $N=1033, 1035$, and 1036, we obtained $\omega_{\rm max}=$ 29.05, 29.08, and 29.08, which are close to the value of $\omega_{\rm max}^{\rm fcc}$, while at $N=1034$, $\omega_{\rm max}=$ 36.83, which is much higher than $\omega_{\rm max}^{\rm fcc}$. 

%%%%%%%%%%%%%%%%%%%%%%%%%%%%%%%%%%%
\section{Conclusion}
In conclusion, we studied the $N$-dependence of the total energy $E$ and the maximum frequency $\omega_{\rm max}$ of the LJ clusters with the size up to $N=1610$. The $\omega_{\rm max}$ reflects the atomic vibrations localized at the core region, and the magnitude of $\omega_{\rm max}$ is significantly small when the core geometry is different from an icosahedron (e.g., $N=38$, 75-77, 98, and 102-104). The $\omega_{\rm max}$ also reflects the vacancy formation at the core and the structural transition from icosahedra to decahedra for large $N$. Based on the relationship between the $E$ and $\omega_{\rm max}$ for the cases of $19\le N\le 39$, we created the metastability map that can provide both transformation pathways between different structures and an estimation of the energy barrier height. 

We hope that the $\omega_{\rm max}$-based approach is applied to study the metastability of more realistic systems including metallic and semiconducting clusters by using accurate potentials. On the other hand, the allotropes of fullerene molecules will show different $N$-dependence of $\omega_{\rm max}$ because of the hollow spherical structures.

%%%%%%%%%%%%%%%%%%%%%%%%%%%%%%%%%%%
\begin{acknowledgments}
This work was supported by JSPS KAKENHI (Grant No. JP21K04628). The computation was carried out using the facilities of the Supercomputer Center, the Institute for Solid State Physics, the University of Tokyo, and using the supercomputer ``Flow'' at Information Technology Center, Nagoya University.
\end{acknowledgments}

%%%%%%%%%%%%%%%%%

%\section*{Data Availability Statement}
%The data that support the findings of this study are available from the corresponding author upon reasonable request.

%%%%%%%%%%%%%%%%%
%\appendix
%\section{Dispersion curves of 2D triangular lattice}
%\label{app}

%===================================================================%
%   References
%===================================================================%

\end{document}